\documentstyle[11pt,paspconf,psfig]{article}

\def\etal{{\it et~al.}}
\def\h0{{$H_0$}}

\def\per#1{{~#1$^{-1}$}}

\def\Sub#1{_{_{#1}}}
\def\dm{$\mu\Sub{0}$}
\def\Dm{\mu\Sub{0}}

\def\imm#1{\raise0.4pt\hbox{$\langle$}$#1$\raise0.4pt\hbox{$\rangle$}}
\def\Imm#1{{\raise0.4pt\hbox{$\langle$}{#1}\raise0.4pt\hbox{$\rangle$}}}
\def\Square#1{
{\raise0.4pt\hbox{$[$}{#1}\raise0.4pt\hbox{$]$}}}
\def\log#1{${\rm log}(#1)$}
\def\Log#1{{\rm log}(#1)}

\def\lsim{{\small\mathrel{\hbox{\rlap{\hbox{\lower2pt\hbox{$\sim$}}}\raise2pt\hbox{$<$}}}}}
\def\gsim{{\small\mathrel{\hbox{\rlap{\hbox{\lower2pt\hbox{$\sim$}}}\raise2pt\hbox{$>$}}}}}
\def\ie{{\it ie.}}
\def\eg{{\it eg.}}
\def\cf{{\it cf.}}

\def\wpl{{$W$-\log{P}}}

\begin{document}

\title{Cepheid Standard Candles}
\author{N. R. Tanvir}
\affil{University of Cambridge, Institute of Astronomy, Madingley Road,
Cambridge, CB3 0HA. United Kingdom.}

\begin{abstract}
Thanks  to HST, there are now many galaxies with 
Cepheid distances and these provide the main platform for
the calibration of
the secondary distance indicators.
I review recent progress in our understanding of the
standard candle properties of Cepheids with particular
emphasis on the techniques used in the HST studies.
The PL relation defined by Cepheids in the LMC is shown
to be excellent, although the distance to the LMC, which
determines the zero-point, is still rather controversial.
This LMC relation is consistent with the galactic calibrations
which use Hipparcos parallaxes or Baade-Wesselink distances.
However, the PL plot for Cepheids observed in galactic open-clusters
is suggestive of an age dependence of
the main-sequence fitting distances,
similar to 
that seen for clusters with Hipparcos parallax distances.
Observational and theoretical studies suggest that 
the metallicity dependence of Cepheid properties is not
large, but is sufficiently important that it should be
accounted for.
However,  
the fact that
target galaxies are typically of similar metallicities to the
calibrators, suggest that metallicity corrections won't have
a major impact on estimates of the Hubble constant.
Incompleteness biases can also affect Cepheid samples, but
are usually best dealt with by imposing a conservative lower
limit on period.
\end{abstract}


\section{Introduction}

A ``ladder'' long ceased to be a good analogy for the cosmic
distance scale since there is always an overlap  of
well-established distance indicators
spanning any particular regime in distance.
But, whatever picture one chooses,
the fact remains that Cepheid variables 
are, as they were in Hubble's time, the most
important
primary distance indicators, used to provide
the step from our galaxy to the nearby universe.
Amongst the ``desirable properties'' of Cepheids
(used here and throughout, unless otherwise indicated, 
to mean classical population-I
Cepheids pulsating in the fundamental mode) are that they (a) are
bright, compared to most other stellar distance
indicators (see figure 1), and easy to 
recognise by their variability; (b) have
been studied long and hard   and are physically pretty well
understood; (c) are long-lived and stable,
and hence can be reobserved;
(d) individually are precise distance indicators which do not rely
on the integrated properties of a whole population; and (e)
are common enough in late-type galaxies that
large samples can be accumulated to beat down statistical
noise.
On the other side of the coin, of course, classical Cepheids (i)
cannot with current technology be observed far enough into
the Hubble flow to give \h0\ directly; (ii) are population I objects,
so are not found in early-type galaxies; (iii) in the
Milky-Way are barely within the reach of geometrical distance
determination and tend to suffer high extinction due to dust
in the disk; and (iv) are still difficult to model in some 
respects, particularly concerning 
the location of the red edge of the instability
strip in the Hertzsprung--Russell diagram.
The questions I seek to address here are, just how reliable are the 
Cepheid based distance
indicators, and are they in harmony with other
indicators with which they can be compared?

\begin{figure}
\centerline{
\psfig{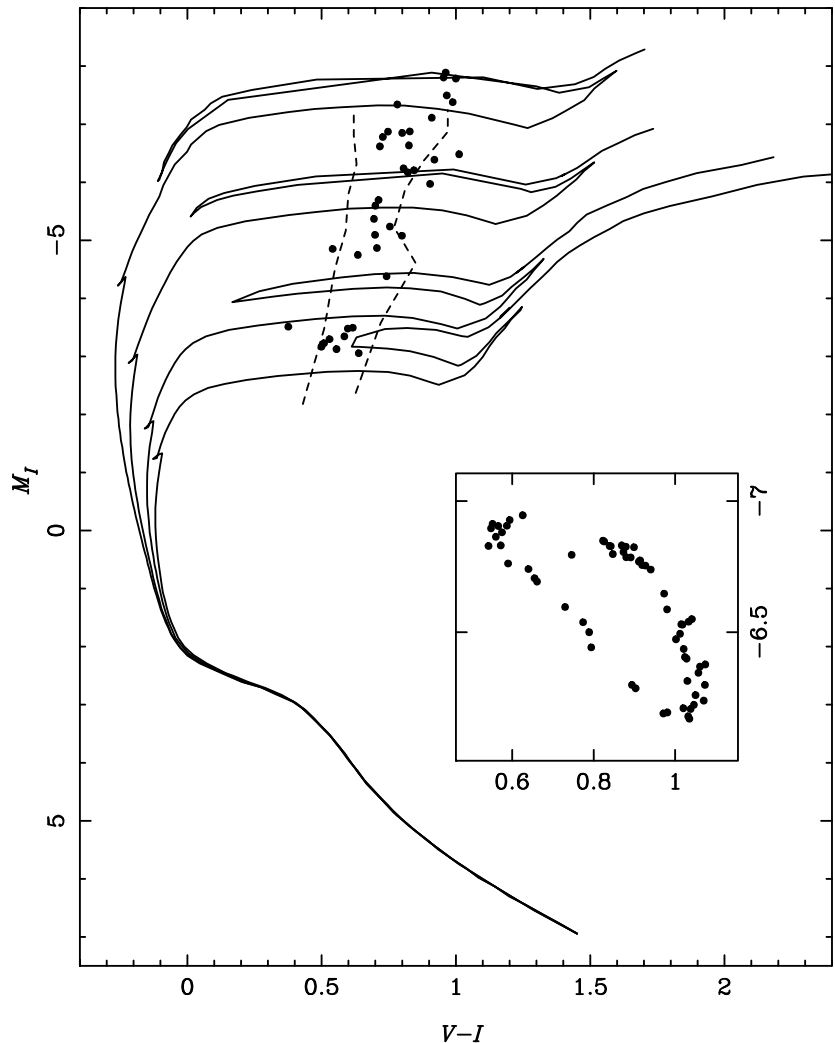}
}
\caption{This colour--magnitude diagram shows how the
well-sampled
Cepheids in the LMC populate the instability strip.
The location of the strip given in the theoretical study
of Chiosi \etal\ (1993; note,
loci are actually for their mass-luminosity relation c and chemical
composition $Y=0.25$, $Z=0.008$) is shown by the dashed lines.
Solid lines are representative isochrones (kindly supplied by
Guy Worthey and based on the Bertelli \etal\ 1994 isochrones)
with ages 16, 40, 100 and 160 Myrs. Most Cepheids are thought to
be in the process of their second crossing of the CM
diagram.
The variation of colour and magnitude around a cycle is 
plotted in the inset panel for one particular well
sampled Cepheid (HV2257 from Moffett \etal\ 1998).
}
\end{figure}

In practice, the Cepheid PL relations are calibrated locally
in the Milky-Way and Magellanic Clouds.
Prior to HST, Cepheids had only been well studied in local
group and other galaxies within about 4 Mpc (Cepheids had been
detected in more distant galaxies, but only in very small numbers
with few epochs).  An extensive summary can be found
in Jacoby \etal\ (1992; their table 1), from which it
can be seen that at that time very few of these galaxies
were useful for checking and calibrating secondary distance indicators
such as the Tully-Fisher relation.
With HST the situation has changed dramatically so that
approaching 30 new galaxies (see table 1 for those
published to date)
have been observed for Cepheids
reaching to  30 Mpc or more.  Furthermore, these were
mostly chosen specifically to be useful
from the point of view of secondary indicators.

\begin{table}
\begin{center}
\caption{Cepheid distances to HST observed galaxies thus far reported.
The values are taken directly from the papers, although the
methodology varies slightly in some cases. Note that where two
errors are quoted the first represents the random component and
the second is an estimate of the systematic component.}
\begin{tabular}{lcl}
\tableline
Galaxy & Published distance modulus & Reference \\
\tableline
M81\tablenotemark{a}     & $27.80\pm0.19$ & Freedman \etal\ 1994\\
M95\tablenotemark{a}     & $30.01\pm0.19$ & Graham \etal\ 1997\\
M100\tablenotemark{a}    & $31.03\pm0.17$ & Ferrarese \etal\ 1996\\
M101\tablenotemark{a}    & $29.35\pm0.17$ & Kelson \etal\ 1996\\
NGC925\tablenotemark{a}  & $29.84\pm0.16$ & Silberman \etal\ 1996\\
NGC1365\tablenotemark{a} & $31.43\pm0.20\pm0.18$ & Silbermann \etal\ \apj\ in press\\
NGC2090\tablenotemark{a} & $30.45\pm0.16\pm0.16$ & Phelps \etal\ 1998\\
NGC2541\tablenotemark{a} & $30.47\pm0.11\pm0.12$ & Ferrarese \etal\ 1998\\
NGC3621\tablenotemark{a} & $29.13\pm0.18$ & Rawson \etal\ 1997\\
NGC4414\tablenotemark{a} & $31.41\pm0.17\pm0.16$ & Turner \etal\ 1998\\
NGC4725\tablenotemark{a} & $30.50\pm0.16\pm0.17$ & Gibson \etal\ \apj\ in press\\
NGC7331\tablenotemark{a} & $30.89\pm0.14$ & Hughes \etal\ 1998\\
NGC4639\tablenotemark{b} & $32.03\pm0.22$ & Saha \etal\ 1997\\
NGC4496A\tablenotemark{b} & $31.03\pm0.14$ & Saha \etal\ 1996a \\
NGC4536\tablenotemark{b} & $31.05\pm0.13$ & Saha \etal\ 1996b \\
NGC5253\tablenotemark{b} & $28.10$ &  Saha \etal\ 1995 \\
IC4182\tablenotemark{b}  & $28.36\pm0.09$ &  Saha \etal\ 1994 \\
M96         & $30.32\pm0.16$ & Tanvir \etal\ 1995\\
\tableline\tableline
\end{tabular}
\end{center}
\tablenotetext{a}{observed by the distance scale key-project team.}
\tablenotetext{b}{observed by the Sandage \etal\ SNIa calibration project.}
\end{table}

Having so many galaxies observed with the same instrument,
and indeed essentially the same procedure and calibrations,
is good for consistency, but makes the whole distance scale
more vulnerable to shared {\em systematic} errors.
Therefore it is most important to address the specific
procedures used in the HST studies and to concentrate
calibration efforts accordingly.

\section{The ``HST Method''}

The standard strategy adopted by all groups using the HST
to observe Cepheids owes much to the ideas developed
particularly by Madore and Freedman (1991, and references 
therein).
Typically, the field is monitored in the $V$-band
at 12 to 15 epochs, to identify variables and
determine periods, phases and
$V$-band amplitudes and magnitudes.
At  3 to 5 epochs $I$-band observations are also
obtained to provide colours and hence a handle on
the reddening.  The lower amplitude in $I$ and
the correlations between the shapes of the $V$ and $I$
light curves mean that the smaller number of epochs is
adequate.

At this point most studies have proceeded to fit $V$- and
$I$-band period-luminosity relations independently
to both sets of magnitudes to obtain apparent
distance moduli in each band.  The difference is assumed
to be due to reddening and hence an unreddened distance
modulus is estimated.
An alternative, but essentially equivalent, procedure is
to calculate reddening-free Wesenheit indices for
each Cepheid and fit an appropriate relation to these
(see T97 for further details).
For $VI$ photometry, the index is defined (\cf\ Madore 1982) as:

$$W\Sub{VI}=\Imm{V}-R\Square{\Imm{V}-\Imm{I}}$$

which is explicitly independent of extinction if:

$$R=\frac{A\Sub{V}}{A\Sub{V}-A\Sub{I}}$$

$R$ is conventionally taken to have a value of 2.45
based on the extinction curve of Cardelli \etal\ (1989).

An advantage of the Wesenheit approach is that it
reveals graphically how the strongly correlated residuals in
both bands (the period--luminosity--colour relation)
are such that the ``reddening corrected'' relation
is intrinsically tighter than the
PL relations in either band individually, even in the
absence of any extinction (\eg\ see next section).
This is not to say that the \wpl\ relation is equivalent
to a period--luminosity--colour relation, but we are
accounting for at least part of the intrinsic colour term.

There are, of course, other ways in which Cepheids
are used as distance indicators, such as using multicolour
(\eg\ Madore \& Freedman 1991; Martin, Warren \& Feast 1979)
and/or infrared photometry(\eg\ Laney \& Stobie 1994), but
this ``HST method'' has the advantage that it 
provides good Cepheid distances
for a  comparatively small expenditure of telescope time.

\section{Calibration via the LMC Cepheids}

The LMC is thought to have 
little depth along the line of sight, comparatively
low extinction and is rich in Cepheids. 
Thus it is a good place to study Cepheid properties and, in recent years,
extragalactic studies have mostly used PL relations 
derived in the LMC.  To calibrate these relations requires,
in the first place, observations
of a good sample of its Cepheids in the relevant bands.
To address this, I have collected all the
published data for Cepheids with Johnson $V$-band and Cousins $I$-band
photoelectric photometry (Tanvir in preparation).  

The PL relations are shown in
figure 1.  
The \wpl\ relation is linear and is at least as
tight as the infra-red PL relations (\cf\ Laney \& Stobie 1994).
The dispersion of 0.12 mag
is  remarkably small  given that the effects of
measurement and sampling errors, aswell as the depth within
the LMC, must be present in addition to the intrinsic spread.
The linear fit to the \wpl\ relation, referred to a pivot
$\Log{P}$ of 1.4 which is typical for extragalactic samples, is:

\bigskip

\noindent
\fbox{
$M\Sub{W}=-3.411 (\pm0.036) \Square{\Log{P}-1.4} + 11.276 (\pm0.017)$\qquad ;\quad$\sigma_{rms}=0.120$ 
}

\bigskip

To establish the absolute zero-point we must subtract from this
the true, extinction-free distance modulus of the LMC.  Several
other contributions to this proceedings address this important point,
and it is beyond the scope of the present paper to review 
the many available estimates of the LMC distance (see Walker
1998 for such a review). 
Instead I shall list some of the recent estimates
based on ``direct'' methods, which in fact give a good
indication of the range of disagreement (table 2).

\begin{table}
\caption{Results for ``direct'' distance determinations to the LMC}
\begin{center}
{\small
\begin{tabular}{llr} \tableline
Method & Source & \dm(LMC) \\ \tableline
Light echo times for SN1987A & Gould \& Uza (1998) & $<18.37\pm0.04$ \\
Light echo times for SN1987A & Panagia  (1998) & $18.58\pm0.05$ \\
Eclipsing binary HV2274 & Guinan \etal\ (1998) & $18.30\pm0.07$ \\
Multimode RR Lyraes & Alcock \etal\ (1997) & $18.48\pm0.19$ \\
Expanding photosphere of SN1987A & Eastman \& Kirshner (1989) & $18.45\pm0.25$ \\
\tableline\tableline
\end{tabular}}
\end{center}
\end{table}

Estimates which are based on 
Cepheids themselves usually (\eg\ Feast \& Catchpole
1997), but not
always (\eg\ Luri \etal\ 1998), fall at the high
end of this range, whilst estimates using RR Lyraes usually
(\eg\ Luri \etal\ 1998), but not always (\eg\ Reid 1997), fall at the low end.
Clearly there is not yet a concensus about the distance to the LMC
at the 20\% (full range) level, and here I continue to adopt
the working value and $1\sigma$ error
recommended by Madore and Freedman (1991)
of $\Dm=18.5\pm0.1$.

\begin{figure}
\centerline{
\psfig{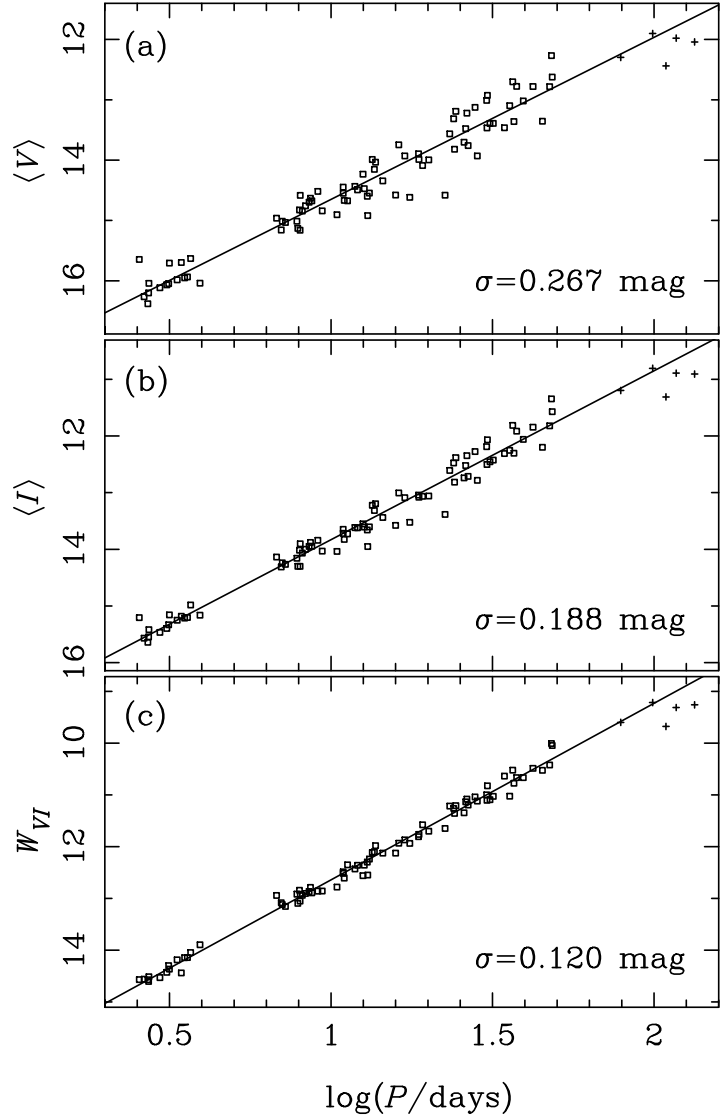}
}
\caption{Period--luminosity relations for LMC Cepheids in
the Johnson $V$-band, Cousins $I$-band and for the
reddening free Wesenheit indices, $W\Sub{VI}$.
The straight-line fit is to the 82 variables with 
${\rm log}(P)<1.8$.
For the numerous sources of data see the references in T97,
to which have been added data from Moffett \etal\ (1998)
and Tanvir \& Boyle (\mnras\ submitted).
}
\end{figure}

In fact, a number of the variables in figure 1 have
as few as 2 observations in each band.  It is interesting
to plot the same relations for only those Cepheids
with well sampled light curves, chosen here to be those
with at least 15 $V$-band observations and 10 $I$-band 
observations (figure 3).  In $V$ and $I$ the
improvement is considerable since sparse sampling of light
curves with upwards of 1 mag peak-to-peak variations, produces
large errors.
However, the improvement in $W\Sub{VI}$.
is less pronounced
because variations in colour around a pulsation
cycle also mimic the effect of dust, in the sense that
reddest colour occurs close to the faintest magnitude.
The increased scatter to shorter wavelengths is produced
in part by differential reddening, but also reflects
the intrinsic width of the instability strip (\ie\ 
variations in temperature at a fixed period have
least effect on the Rayleigh-Jeans tail of the spectrum).

\begin{figure}
\centerline{
\psfig{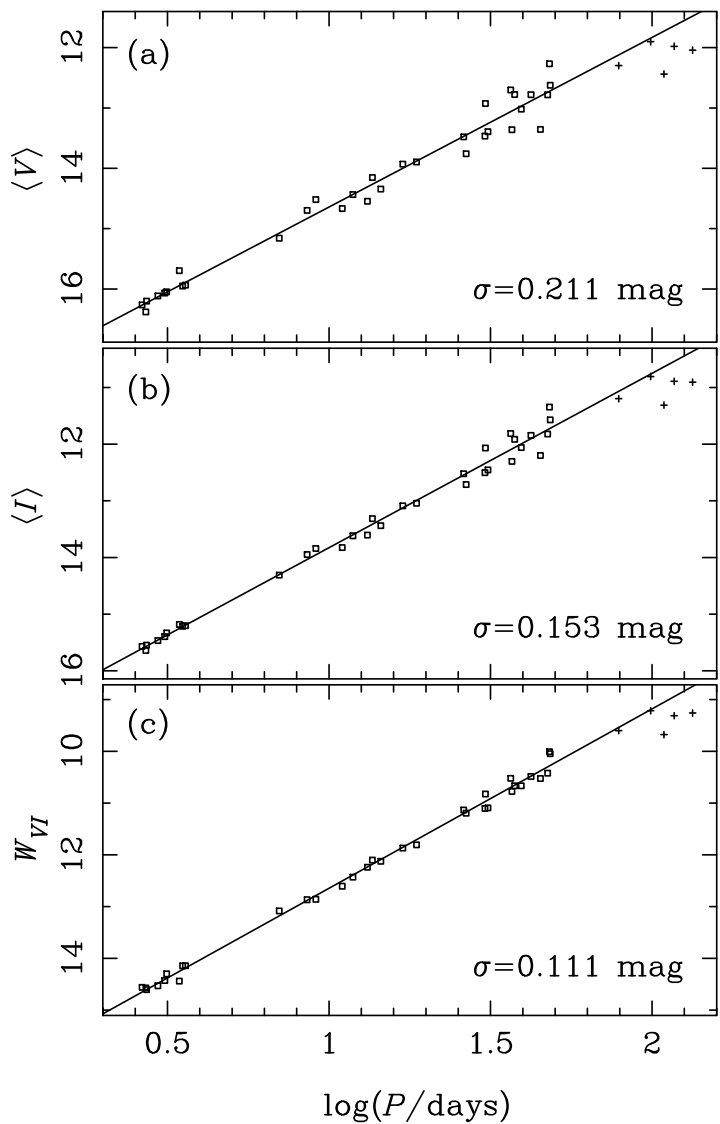}
}
\caption{
As figure 2, but just for the
subset of 33 Cepheids ($\Log{P}<1.8$) which have at least 15 data 
points in
the $V$-band and 10 in the $I$-band.
}
\end{figure}

Since these data are certainly better for $V$ and $I$, 
they are the most suitable for calibrating the PL relations
in those bands:

\bigskip

\noindent
\fbox{
\shortstack{
$M\Sub{V}=-2.810 (\pm0.082)\Square{\Log{P}-1.4} + 13.517 (\pm0.043)$\qquad ;\quad$\sigma_{rms}=0.211$ \\
$M\Sub{I}\ =-3.078 (\pm0.059)\Square{\Log{P}-1.4} + 12.595 (\pm0.031)$\qquad  ;\quad$\sigma_{rms}=0.153$
}
}

\bigskip

\section{Calibration via the Milky-Way Cepheids}

If we can estimate distances to individual Milky-Way
Cepheids, then we can use them to calibrate at least the
zero-point of the period--luminosity
relations.
Here I consider three methods and, 
in the spirit of our quest for harmony,
will check them for consistency with the \wpl\ calibration already derived
from the LMC Cepheids.
For the present, no metallicity corrections are made, but these
are discussed in section 6.

Firstly, although very few individual Cepheids have statistically significant
parallaxes from Hipparcos, it is possible to average large samples
of poorly determined parallaxes  to yield useful, unbiased
calibrations of the PL relations (\eg\ Feast \& Catchpole 1997).
If we define a ``photometric parallax'' for each Cepheid:

$$
\pi\Sub{W}/\gamma=10^{0.2\left(\alpha\Sub{W}\Log{P}-W\Sub{VI}-5\right)}
$$

where $\alpha\Sub{W}$ is the slope of the \wpl\ relation for the LMC sample,
then we can compare to the observed trigonometric parallaxes to
obtain the zero-point. This is plotted in figure 4, which clearly
shows the LMC calibration to be consistent with the data.
We should, of course, beware that when dealing with high extinctions
as is the case for many of these Cepheids, the reddening correction
procedure itself may introduce significant errors.

\begin{figure}
\centerline{
\psfig{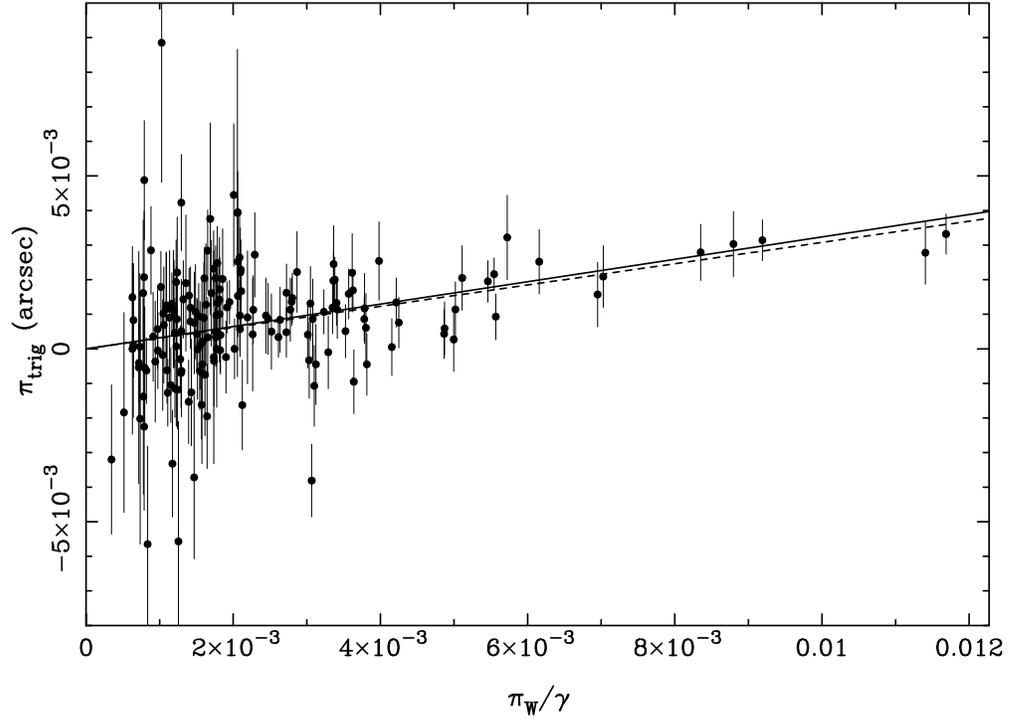}
}
\caption{
Hipparcos trigonometric parallaxes for 177 
Cepheids are plotted against their ``photometric
parallax'' based on $W\Sub{VI}$ (see text).
Both $V$ and $I$ photometry is from the compilation 
of Caldwell \& Coulson (1987).
The slope of the $W\Sub{VI}$-period relation is assumed
from the LMC, and the solid line illustrates the LMC
zero-point.
For comparison, the zero-point obtained by a weighted fit to the
data, constrained to pass through the origin,
is plotted as a dashed line.
The difference ammounts to the LMC calibration being fainter
by 0.11 mag.}
\end{figure}

Secondly, I consider Cepheids with distances found
via Baade-Wesselink methods.  Figure 5 uses
the data from Gieren \etal\ (1997) to which the reader is
referred for a fuller discussion of this important technique.
For our purposes we simply note that the calibrations are
in good agreement, particularly over the range of most
interest for extragalactic studies, namely $\Log{P}>1$

\begin{figure}
\centerline{
\psfig{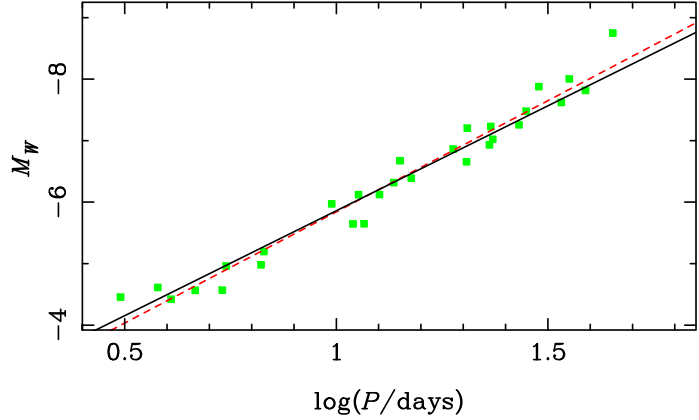}
}
\caption{
Plot of absolute $W\Sub{VI}$ versus period for a sample of Milky-Way
Cepheids with Baade--Wesselink determined distances from
Gieren \etal\ (1997).
The solid line is the fit to the LMC Cepheids from figure 2
and the dashed line is the best fit to this data.
Note that Gieren \etal\ caution against the use of EV Sct and
QZ Nor, the lowest period variables, 
as possible overtone pulsators (also SZ Tau which has indeed
been omitted from this diagram because of its particularly uncertain
status \eg\ see Turner 1992).
They are also unhappy with the technique, as it stands, for
variables with $\Log{P}>1.6$, which is only one point in this
figure.
}
\end{figure}

Finally, I look at the time-honoured method which uses
Cepheids in open clusters with main-sequence fitted distances.
These are plotted in figure 6 and we see immediately that
the galactic Cepheids  define a steeper relation
than the LMC.  
Although apparently a significant difference, we should be cautious 
since the numbers are small and many of the points can be questioned
on an individual basis, for example, as to the reliability of
the association between cluster and Cepheid.

However, if the effect is real, we can ask whether there are
any plausible explanations.
The global metallicity difference between the LMC and Milky-Way
would be a surprising cause since, if anything, increasing
metallicity is expected to produce a somewhat shallower
slope
(\eg\ Bono \etal\ Ap.J. in press, Chiosi \etal\ 1993).
Nonetheless, metallicities
of individual Cepheids
do correlate well with residual for a subset of
this sample with high quality measurements
(Sekiguchi \& Fukugita 1998; Fry \& Carney 1997).  However,
the nearly 
one-to-one correlations of residuals in different passbands combined with
the absence of any very obvious correlation for the BW 
distances (Tanvir 1998) suggest that any such problem would have to be 
largely with the MS fitting distances rather than the Cepheids
themselves.
A simple metallicity effect of this kind has not been seen in the
Hipparcos results.

An alternative, if at first sight even less palatable,
possibility is that the main-sequence fits to the clusters
are dependent on the age of the clusters (assumed to be
the same as the age of the Cepheid which in turn is a function of period). 
The correlation of age and \wpl\ residual (figure 7)
is actually quite good given the various errors and
assumptions, and 
intriguingly the nature of
the effect is similar to that found by van Leeuwen (this volume)
for the age dependence of main-sequence position found by
Hipparcos.
Because of the age--metallicity relation of the galactic
disk (Edvardsson \etal\ 1993), an age correlation should
also be reflected in a metallicity correlation at some level,
but the apparent tightness of the latter (Sekiguchi \& Fukugita 1998)
remains surprising.

Of course, extinction corrections for these Cepheids
are high and difficult to measure
(Hoyle \etal\ in this volume) and this could also be 
affected if the MS fitting is age dependent.

\begin{figure}
\centerline{
\psfig{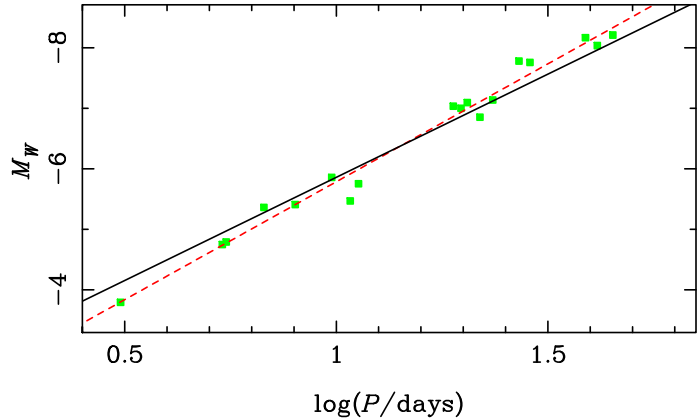}
}
\caption{
Plot of absolute $W\Sub{VI}$ versus period for a sample of Milky-Way
Cepheids in open-clusters or associations with main-sequence fitting 
distances.
The LMC relation is plotted as a solid line and is compared to the
best fit to this data (dashed line).
The cluster distances are taken from the compilation
of Laney \& Stobie (1994), also Turner, Pedreros \& Walker (1998) and
refs therein. Photometry from Caldwell \&
Coulson (1987).
}
\end{figure}

\begin{figure}
\centerline{
\psfig{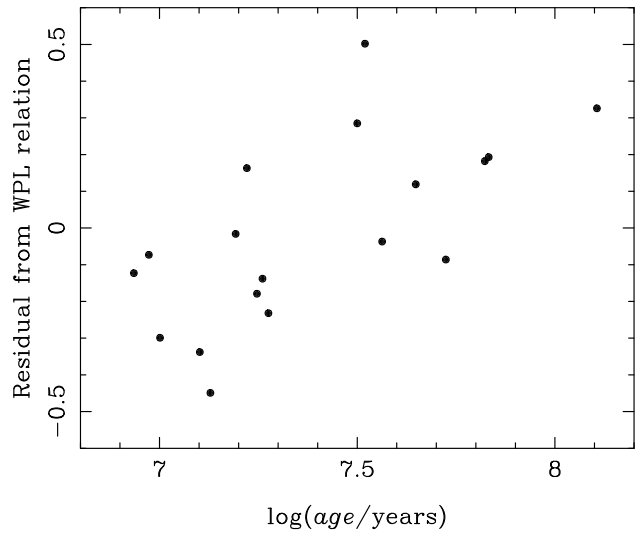}
}
\caption{
Residuals for the Milky-Way open-cluster Cepheids 
from the LMC \wpl\ relation (\ie\ from the solid
line in figure 6) plotted against estimated age of Cepheid.
This is suggestive that the distances estimated by main-sequence
fitting are systematically affected by the age of the cluster
in question, although other explanations are possible.
The ages are found using the relation between turn-off mass
given by Turner (1996) and the theoretical isochrones of 
Bertelli \etal\ (1994).
}
\end{figure}

\section{Deriving Cepheid parameters from sparse, noisy data}

We have seen that surprisingly good Wesenheit indices
are found for the LMC Cepheids with very few epochs of
observation.  However, for faint extragalactic Cepheids
the presence of much greater photometric noise makes
estimating their parameters, both period and magnitudes,
a trickier business.

As described in section 2, the $V$-band data is generally
used to identify variables, find their periods and to determine
at some level the shape of the light curves.
The simplest way to map from the $V$ to the $I$ light curves
is to just scale their amplitudes, which are normally in a  ratio
of about 1:0.6 (T97).
Some more sophisticated variations on this theme:
allow for small shifts in phase (Freedman 1988);
use empirical mappings which are a variable function of phase
(Labhardt, Sandage \& Tammann 1997); or,
use light curve templates derived from fourier fitting to well sampled,
low noise data (Stetson 1996).

A new technique (Tanvir, Hendry \& Kanbur in preparation; see also
Hendry \etal\ in this volume) uses  principal components 
to characterise the light curve shapes of Cepheids as a function
of period.
The correlations between
the light curves in different bands are built-in in this
method, and the
fits are therefore to the $V$ and $I$ data points simultaneously.
This allows all the data to be used in the determination of
the period and magnitudes and uses our full knowledge of the
properties of well-observed, local Cepheids.  It is also
more amenable to the ascription of error bounds on the resultant
parameters.
An example of the method in action is shown in figure 8.

\begin{figure}
\centerline{
\psfig{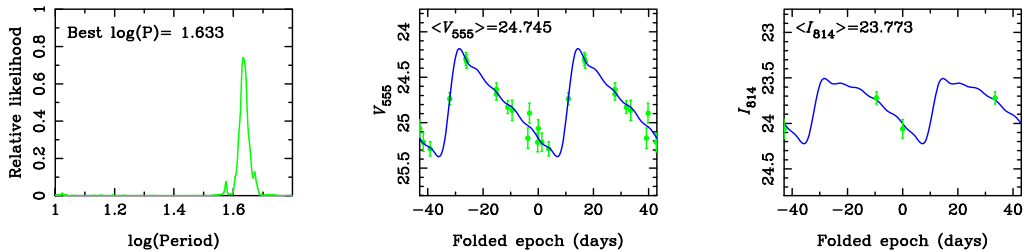}
}
\caption{
An example of the use of light curve templates fitted to 
photometry for a Cepheid in M96.  The procedure finds the
best period and magnitudes of the Cepheid.
}
\end{figure}

\section{The effect of metallicity differences}

One of the long-standing concerns over the use of Cepheids is
the question of what effect chemical abundance variations may have on the
PL relations, and in particular on the \wpl\ relation.
Observationally this is a difficult question to answer because 
samples of Cepheids which are all known to be at the same distance
(\ie\ usually because they are all in one galaxy) tend to have
little variation in metallicity.
Before seeing how recent observational tests are beginning to 
provide useful constraints, we note  that (a) variations
within the LMC sample are apparently not enough to introduce
any significant scatter in the \wpl\ relation (section 3); 
and (b) calibrating Cepheids in the 
LMC and Milky-Way straddle the average metallicity
of the sample of HST observed galaxies, so abundance errors
will tend to cancel out (figure 9).

\begin{figure}
\centerline{
\psfig{figure=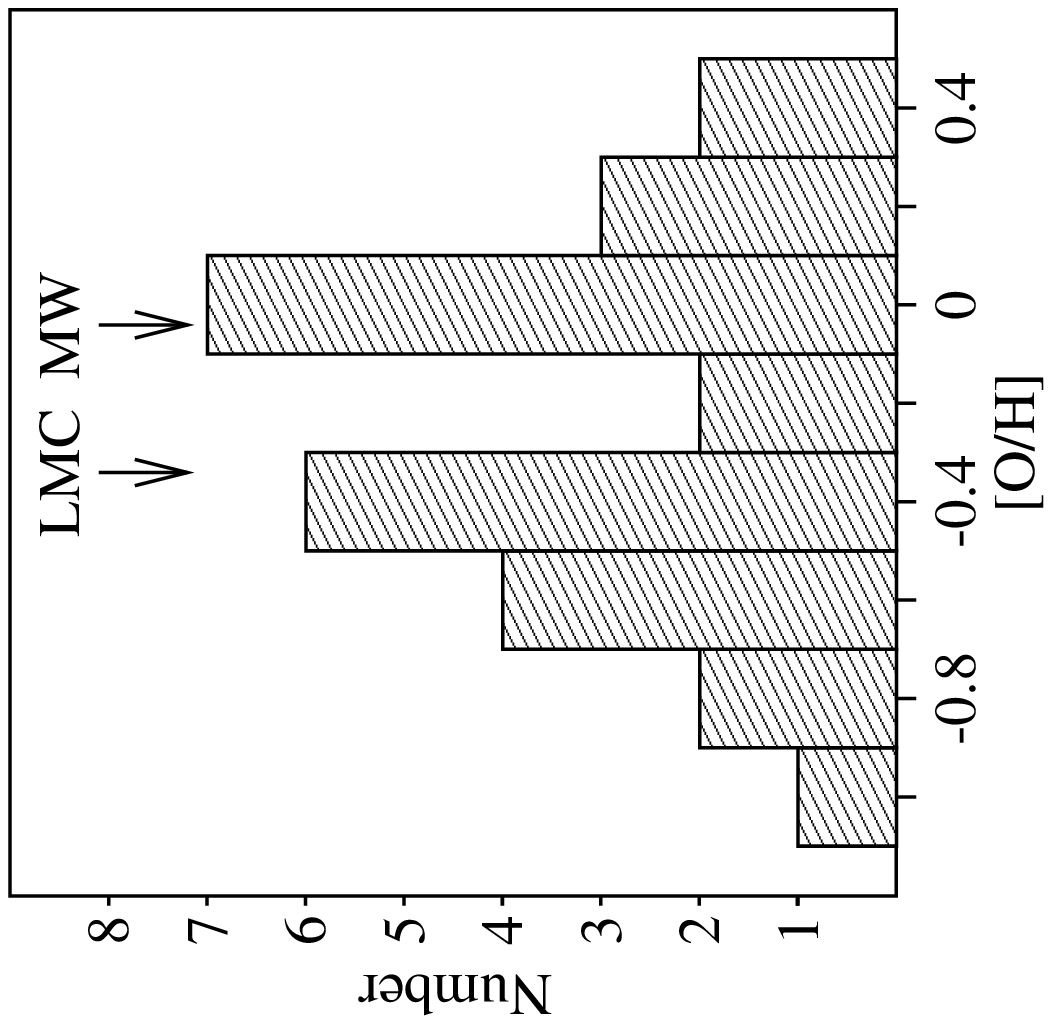,width=7cm,angle=270}
}
\caption{
Metallicities for the 21 key-project
target galaxies and another 6 galaxies which have been
observed by HST for other projects. 
On average, the LMC and MW calibrations are not likely to
produce results for \h0\ which are significantly
biased as a result of metallicity variations.
Data is from the compilation by Kennicutt \etal\ (1998;
who show a  similar figure)
and Kochanek (1997).
}
\end{figure}

There have been a number of attempts over the years to tie down
the metallicity dependence of Cepheid properties observationally.
All have rather high formal uncertainties and some are not really
applicable to the ``HST method'' in that they are concerned with
other passbands.
Several recent, relevant efforts are summarized in table 3.
From these it appears that the difference in metallicity
between the Milky-Way and LMC Cepheids (about 0.3 in $[Fe/H]$)
should only lead to a $\sim0.1$ mag difference in zero point.

\begin{table}
\caption{Recent results for the metallicity sensitivity
of Cepheid distances where the distance determination uses
the \wpl\ relation to correct for reddening on the basis of
the Cepheids themselves. The sense is that a target sample
whose metallicity is {\it higher} than the calibrator sample
will be found to have a spuriously {\it low} distance modulus, if
the value in column 1 is negative.}
\begin{center}
\begin{tabular}{lll} \tableline
$\delta\mu_0/\delta{[O/H]}$ & Method & Reference \\
(mag\per{dex}) & & \\
\tableline
$-0.24\pm0.16$ & 
\parbox[t]{5.5cm} {Comparison of HST observations of inner and outer fields of M101.} 
& Kennicutt \etal\ (1998)\\
& & \\
$-0.4\pm0.2$ & 
\parbox[t]{5.5cm}{Simultaneous solution for distances to 17 galaxies} 
& Kochanek (1997).\\
& & \\
$-0.44_{-0.2}^{+0.1}$ & 
\parbox[t]{5.5cm}{Comparison of EROS observations of SMC and LMC Cepheids} 
& Sasselov \etal\ (1997).\\
\tableline\tableline
\end{tabular}
\end{center}
\end{table}

On the theoretical side, there are a number of
obstacles to providing exact 
predictions of PL and PLC relations and their metallicity
dependence.  These include locating the position of the
red-edge of the instability strip, which is determined by
the onset of convection.
The models of Chiosi, Wood and Capitanio (1993) indicated
a small metallicity dependence for \wpl\ distances (T97).
Recently Bono \etal\ (Ap.J. in press; see also 
Marconi \etal\ in this volume) have developed more
sophisticated, convective pulsation models which actually 
predict a modest metallicity effect in the opposite direction
to that found in the observational studies.
Further progress in theoretical modelling, in parallel with
observational studies, would be very welcome.

\section{Biases due to incomplete samples}

T97 discussed at some length the issue of biases due to
incompleteness of the Cepheid samples.
These arise generally because 
close to
the detection threshold Cepheids preferentially fall into or out of 
the sample depending on whether they happen to be brighter
or fainter than the average PL relation.
From our point of view we are interested not in the simple
$V$-band PL relation, but in the more complicated
\wpl\ relation.
Since, the intrinsic 
dispersion of the \wpl\ relation due to the width of the instability strip
is very small (section 3), the bias will depend largely 
on the other, observational errors, and in particular those which
are largest close to the detection limit.
T97 showed that if these errors are uncorrelated between the
bands (\eg\ if photometric noise dominates) 
then the resulting bias  actually produces spuriously
large distance estimates, which is in the opposite sense
to the normal incompleteness bias (\eg\ Teerikorpi 1987).
This assumes, as is usually the case in practice, that
if a variable is bright enough to be identified as such
in the $V$-band, then it will always be possible to estimate
an $I$-band magnitude for it.  In other words that the selection 
is only on the $V$ magnitudes.

However, if the errors in $V$ and $I$ are correlated, such
as is expected for crowding errors or
uncertainties in the period determination, then the nature of
the resultant bias depends on the details of the correlations.
For example, if the residuals from the $V$ and $I$ PL relations
are essentially correlated one-to-one, then a bias arises which is in the 
traditional sense of an underestimate of the distance.

Lanoix \etal (\apj, submitted) have shown that for a particular
choice of (plausible) random and correlated errors it is possible
to reproduce quite well the apparent bias in the NGC4536 
Cepheid sample.
Unfortunately, to do a good job of estimating the bias for most 
samples
would require simulating Cepheids by adding artificial stars 
to the images.  However, a signature of some kind of bias
is if the Cepheids in a sample systematically
depart from the slope of the fitted \wpl\ relation at short
periods, and the safest action is to
make a conservative lower cut in \log{P}.

\section{Conclusions}

The very tight \wpl\ relation in the LMC illustrates
the considerable power of the ``HST method'' for Cepheid
distance determination.  A dispersion of 0.12 mag, or
less, around the mean relation, implies that a single well
observed Cepheid can, in principal, 
give reddening-free distances to better than 6\% {\it rms}.
To achieve this potential requires: a good calibration,
which would be achieved if we had a definitive (harmonious!)
distance to the LMC;
an understanding of systematic effects, particularly
metallicity; and, of course, good, well calibrated data
(see Hill \etal\ 1998 for a discussion of the issues in
calibrating WFPC2 photometry).
The apparent variation in the location of the main-sequence
seen in the Hipparcos data may yet have profound implications
for the distance scale (and astrophysics), but at least
the evidence to date suggests that the Cepheid calibration
is not too badly affected.

\acknowledgments

I would like to thank my collaborators Martin Hendry, Shashi Kanbur,
Tom Shanks, Fiona Hoyle and Floor van Leeuwen for many useful
discussions.

\end{document}